\documentclass[12pt,preprint]{aastex}
\usepackage{graphicx}
\begin{document}

\slugcomment{Submitted to the Astrophysical Journal}

\title{The mid-infrared spectrum of the transiting exoplanet \\ HD 209458b}

\author{M. R. Swain\altaffilmark{1}, J. Bouwman\altaffilmark{2},
R. L. Akeson\altaffilmark{3}, S. Lawler\altaffilmark{3}, \&
C. A. Beichman\altaffilmark{3} }

\altaffiltext{1}{Jet Propulsion Laboratory, 4800 Oak Grove Drive, Pasadena, CA 91109} 
\altaffiltext{2}{Max-Planck Institute for Astronomy, Konnigstuhl 17, D-69117 Heidelberg Germany} 
\altaffiltext{3}{Michelson Science Center, California Institute of Technology , MS 100-22, Pasadena, CA, 91125}

\begin{abstract}

We report the spectroscopic detection of mid-infrared emission from
the transiting exoplanet HD 209458b.  Using archive data taken with
the Spitzer/IRS instrument, we have determined the spectrum of HD
209458b between 7.46 and 15.25 $\mu$m.  We have used two independent
methods to determine the planet spectrum, one differential in
wavelength and one absolute, and find the results are in good
agreement.  Over much of this spectral range, the planet spectrum is
consistent with featureless thermal emission.  Between 7.5 and 8.5
$\mu$m, we find evidence for an unidentified spectral feature.  If
this spectral modulation is due to absorption, it implies that the
dayside vertical temperature profile of the planetary atmosphere is
not entirely isothermal.  Using the IRS data, we have determined the
broad-band eclipse depth to be 0.00315 $\pm$0.000315, implying
significant redistribution of heat from the dayside to the nightside.
This work required development of improved methods for Spitzer/IRS
data calibration that increase the achievable absolute calibration
precision and dynamic range for observations of bright point sources.

\end{abstract}
\keywords{planetary systems --- stars: individual (HD 209458) ---
methods: data analysis --- techniques: spectroscopic}

\section{Introduction}

The Spitzer Space Telescope has revolutionized the observational
characterization of exoplanets by detecting infrared emission from
these objects; measurements have been reported for HD 209458b
\citep{deming05a}, TrES-1 \citep{charbonneau05}, HD 189733b
\citep{deming06}, and $\upsilon$ Andromeda b \citep{harrington06}.  HD
209458b, the first reported transiting exoplanet
\citep{charbonneau00}, is located at a distance of 47 pc and has a G0
stellar primary (V = 7.6 mag).  The most recent system parameters for
this hot Jovian exoplanet have been established by \cite{knutson07},
with $P = 3.52474859\pm$0.00000038 days, $M_{planet} = (0.64 \pm
0.06) M_{J}$, an eccentricity consistent with zero, and
$R_{P}=(1.320\pm0.025)R_{J}$, 10-20\% larger than predicted by
irradiated planet models.  The detection of infrared emission from hot
Jovian exoplanets has stimulated extensive theoretical work on the
atmospheric structure and emission of these planets.  Constraining the
model predictions for infrared emission from hot Jovian atmospheres is
an important motivation for current observing programs. 

Spectral characterization of hot Jovian exoplanets is a high priority
and is essential for understanding atmospheric composition and
properties.  Spectroscopic detection of exoplanet emission has proved
challenging from the ground \citep{richardson03,deming05b}; space-based
infrared spectroscopy is particularly appealing due to the absence of
an atmosphere, improved signal-to-noise (SNR), and instrument
stability.  Recently, the announcement of a Spitzer/IRS detection of a
featureless emission spectrum from HD 189733b \citep{grillmair07} and
an emission spectrum containing emission features from HD 209458b
\citep{richardson07} has generated great interest.  However,
observations with the Spitzer IRS instrument are complicated by
systematic errors that are large compared to the observable signature.
Some of these systematic errors introduce wavelength-dependent
effects; thus, careful calibration and validation is essential.  In
this paper we present results based on a new approach for calibrating
the major instrument systematic effects affecting these observations.
Using data taken from the Spitzer archive, we have determined the
spectrum of of HD 209458b using two semi-independent methods.

\section{Observations}
\label{observations}

The observations we analyzed (originally proposed by Richardson et
al. 2007) were taken with the Spitzer Space Telescope \citep{werner04}
using the Infrared Spectrograph (IRS; Houck et al. 2004).  The data
were taken on 6 July 2005 and 13 July 2005 as two separate
Astronomical Observing Requests (AORs 14817792 and 14818048) and
provide approximately continuous coverage of the secondary eclipse
event (see Fig. \ref{fig:initialTimeSeries}).  The timing of the
observations is well suited for application of the secondary eclipse
technique (also termed ``occultation spectroscopy"), in which data
from portions of the orbit where light originates from the
``star+planet" and ``star" are subtracted to obtain the planet's
emission \citep{richardson03}.  For both sets of observations, the IRS
instrument was operated in first order (7.5 to 15.2 $\mu$m) at low
spectral resolution (R=60-120; SL1) with a nod executed at the midpoint of
the observations.  This observational sequence provides two completely
independent data sets that span an interval covering the sequence:
\begin{enumerate}
\item before eclipse (flux originates from star+planet),
\item ingress (planet flux contribution changing with time),
\item secondary eclipse (flux originates from star only),
\item egress (planet flux contribution changing with time), and
\item after eclipse (flux originates from star+planet).
\end{enumerate}
\noindent Each nod contains 140 samples with an integration time
of 60 seconds each.

To determine the orbital phase of HD 209458b we used the results by 
\cite{knutson07} for both the period and the ephemeris.  The time for
each data point was determined using the {$\rm DATE\_OBS$} keyword in
the header, which was then converted to Julian date using the IDL
routine JDCNV.pro from the IDL astronomy library. We then converted to
heliocentric Julian date (HJD) using the IDL routine {$\rm
HELIO\_JD.pro$} (also from the astronomical library) for direct
comparison with the \cite{knutson07} results.  The phase was then
estimated by $phase = ((time\_in\_HJD - ephemeris)\times
mod(period))/period$.  In what follows, we will refer to the segment
of the orbital phase when both star and planet are visible as ``SP".
Similarly, we refer to the segment of orbital phase when only the star
is visible (when the planet is passing behind the star) as ``S".  To
determine the planet spectrum, we have applied the analysis described
below to the spectral range of the IRS SL1 module.

\section{Analysis}
\label{analysis}

The extracted flux density time series suffers from four kinds of
temporal changes (see Fig. \ref{fig:initialTimeSeries}) that
completely dominate (by a factor of $\sim$10) the expected signature
of secondary eclipse flux decrement of $\sim$0.0025 \citep{deming05a}.
These effects are (i) a flux offset between nods, (ii) a periodic flux
modulation, (iii) initial flux stabilization, and (iv) monotonic flux
drift within a nod.  These temporal changes are not random; a scatter
diagram shows that the flux density values are highly correlated
(correlation coefficients of $\sim$0.99). We find that these four
major temporal flux density changes listed above are caused by (in
order of importance) errors in telescope pointing, background
subtraction, and latent charge accumulation.

Effective calibration of these systematic effects can be challenging
to demonstrate.  To test our control of the systematic errors, we
developed two methods for estimating the exoplanet spectrum.  The
first method is differential and has the property that errors which
are ``common-mode'' in wavelength are rejected.  The second method is
an absolute method and results in an exoplanet spectrum in units of Jy.  A
schematic picture of our data reduction method is shown in
Fig. \ref{fig:cal_diagram}; central to our approach is comparing the
results of the two semi-independent estimates of the exoplanet
spectrum.  Because the two methods interact with systematic errors
differently, the comparison is useful for accessing the
level of uncalibrated, residual systematics.  In this section, we
describe the initial data extraction, the major systematic errors, and
each of our spectral extraction methods.  In Section 4, we discuss the
comparison between the differential and absolute methods for
obtaining the planet spectrum.  We then present our results and
discuss the implications.  We also discuss the differences between the
methods and results of our approach and previous work.

\subsection{Data Extraction}

Our initial data extraction method is an extension of the method
described by \cite{bouwman06}.  The series of extracted images are
used to define a median background image for each of the two nod
positions.  The median background image (for each nod position) is
then subtracted from all the individual observations with the source
in the other nod position; this generates the background subtracted
images.  We then identify bad pixels using a median filter and visual
inspection.  The bad pixels are then corrected using an approach
similar to the Nagamo-Matsuyama filtering method \citep{nagano79}.
The source spectrum is then extracted using the method developed by
\cite{higdon04} and implemented in the SMART data reduction package.

The spectra were extracted using a fixed-width aperture of 6 pixels
centered on the position of the source.  The exact source position
relative to the slit was determined by fitting a $sinc$ profile to the
spectra in the dispersion direction using the collapsed and normalized source
profile. The accuracy at which the source position can be determined
is about 0.02 pixels. This, together with the aperture width of 6
pixels, ensures that any flux variablity due to slight changes in the
positioning of the aperture are far less than the expected planetary
flux.

\subsection{Systematic Errors}

Here we discuss the origin and chromaticity of the three significant
systematic errors present in these data.  There may be other
systematic errors as well, but they, and the residuals of the errors
we explicitly deal with, are smaller than the uncertainty level
achieved in our calibration.  We acknowledge that there are different
points of view regarding the calibration of IRS data for determining
exoplanet spectra \citep{richardson07,grillmair07} and that these
approaches may perform similar (but not identical) corrections to the
data while ascribing the underlying systematics to different causes.
However, ours is the only method that allows determining the absolute
planet spectrum.

\subsubsection{Pointing Errors} 
The periodic and linear drift components of the Spitzer pointing error
have been documented with long-duration IRAC observations
\citep{morales06}.  Pointing errors cause modulation of the measured
flux because telescope motion perpendicular to the slit axis changes
the position of the stellar image with respect to the spectrometer
entrance slit; this causes changes in the vignetting of the stellar
image.  Even small pointing errors change how the wings of the point
spread function (PSF) are vignetted.  Since the PSF size is
proportional to wavelength, the measured flux changes due to pointing
are wavelength dependent.  In the absence of other effects, the
measured flux density, $S(\lambda)$, is
\begin{equation}
S(\lambda) = {\bf \zeta}(y,\lambda) F(\lambda)
\end{equation}
where $F$ is the ``true'' flux density, $\zeta$ is the
pointing-induced fractional flux density ($\zeta =1$ for no pointing
error), and $y$ is the angular error with respect to the spectrometer
entrance slit center position in units of pixels.  In principle, if
$\zeta$ can be determined, the effects of pointing error can
corrected.

We determined $\zeta(y,\lambda)$, by using the spectral map
observations of IRS calibrators HD 173511 (AOR 13481216) and HR 7341
(AOR 16295168).  The spectral map data consist of a series of pointed
observations in which the star spectrum is measured on a
two-dimensional grid (7x7 and 5x23 positions, respectively, for these
AORs).  For each scan perpendicular to the slit axis, we normalized
the measured spectrum by the spectrum measured at the nominal slit
center position, $\zeta(y) = S_{y}/S_{0}$.  Assuming the slit has a
constant width, we combined the normalized measurements from all the
slit scans.  This resulted in a series of values at each nominal
pointing position perpendicular to the slit axis ($y \in [y_{1},
y_{2}, y_{3}, \ldots]$).  The difference in these values at each
nominal slit scan position reflects a pointing error that can be
corrected for in an iterative process.  We defined a ``template'' by
taking the average value of the points at each pointing offset
position.  The individual slit scan data were then shifted in the
horizontal axis and renormalized to minimize the $\chi^{2}$ value of
the shifted curve with respect to the template.  After all the slit
scans had been shifted and renormalized, a linear interpolation was
done to find revised values for the template function at the nominal
pointing offset positions transverse to the slit axis.  The individual
slit scan data were then shifted and renormalized again for a best fit
to the revised template function.  This process was iterated until
convergence was reached; it resulted in pointing-error-corrected,
slit-scanned data.  We determined $\zeta$ by fitting a cubic-spline at
each wavelength through the shifted and renormalized slit scan
measurements (see Fig. \ref{fig:pointingCorrect} top).

\begin{figure}[h!]
\begin{center}
\epsscale{0.5}
\includegraphics[width=4in,angle=-90]{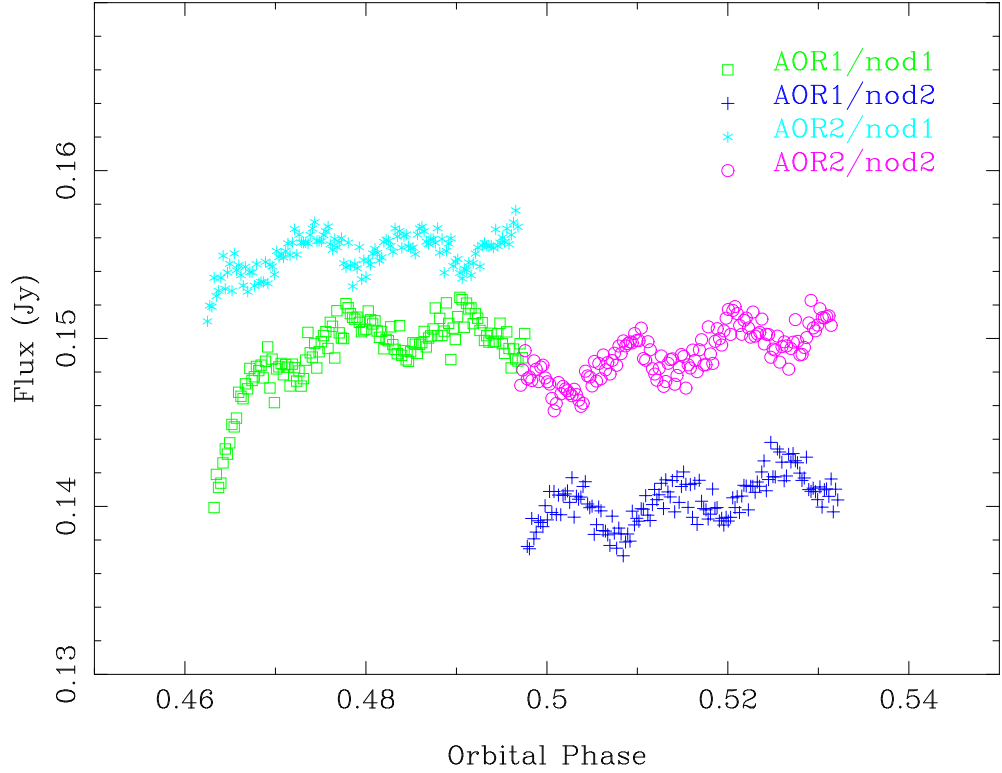}
\caption{The 9 $\mu$m flux time series for AOR 14817792 and AOR
  14818048.  In both AORs, a nod was executed near the center of the
  secondary eclipse period.  The measured flux is affected by
  systematic errors caused by (i) pointing, (ii) background
  subtraction, and (iii) latent charge accumulation; these effects
  combine to cause changes in the measured flux that are $\sim$10x
  larger than the secondary eclipse signature.  The {\bf pointing
    error} causes the offset between the flux density in the nod 1
  data in both AORs, and it causes most of the offset in flux density
  between the nod 1 and nod 2 data for a given AOR; the pointing error
  also causes the linear and periodic changes in the measured flux.  The {\bf
    background subtraction} error causes a portion of the offset
  between the nod 1 and nod 2 data for a given AOR.  The {\bf latent
    charge accumulation} error causes the initial rapid increase in
  measured flux; this effect is larger in AOR 14817792 due to the
  array history.
\label{fig:initialTimeSeries}}
\end{center}
\end{figure}

\begin{figure}[h!]
\begin{center}
\epsscale{0.5}
\includegraphics[width=4in,angle=0]{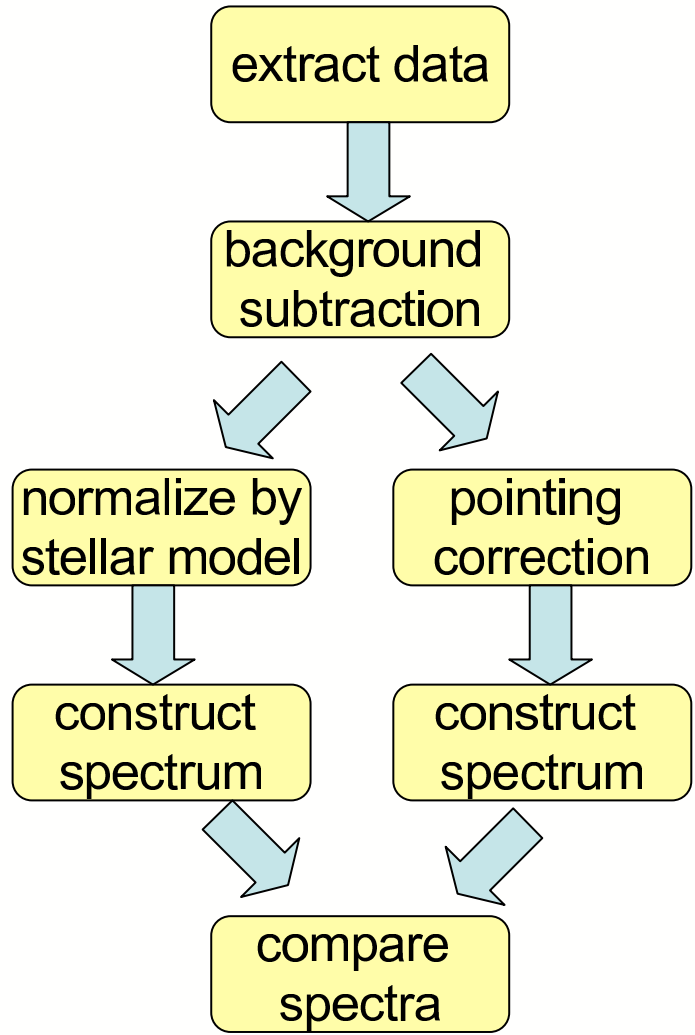}
\caption{A schematic representation of the steps in our data reduction
procedure; a detailed discussion of each stage of this process is
contained in the text.  The strength of this approach is that it
provides two semi-independent methods of estimating the planet
spectrum.  Because the two methods interact with systematic errors
differently, the comparison of the results provides a robust method
for demonstrating control of these errors.  In the final step
of the data reduction, the estimates of the planet spectra from the
two methods are compared.
\label{fig:cal_diagram}}
\end{center}
\end{figure}

\begin{figure}[h!]
\begin{center}
\epsscale{0.5}
\includegraphics[width=4in,angle=0]{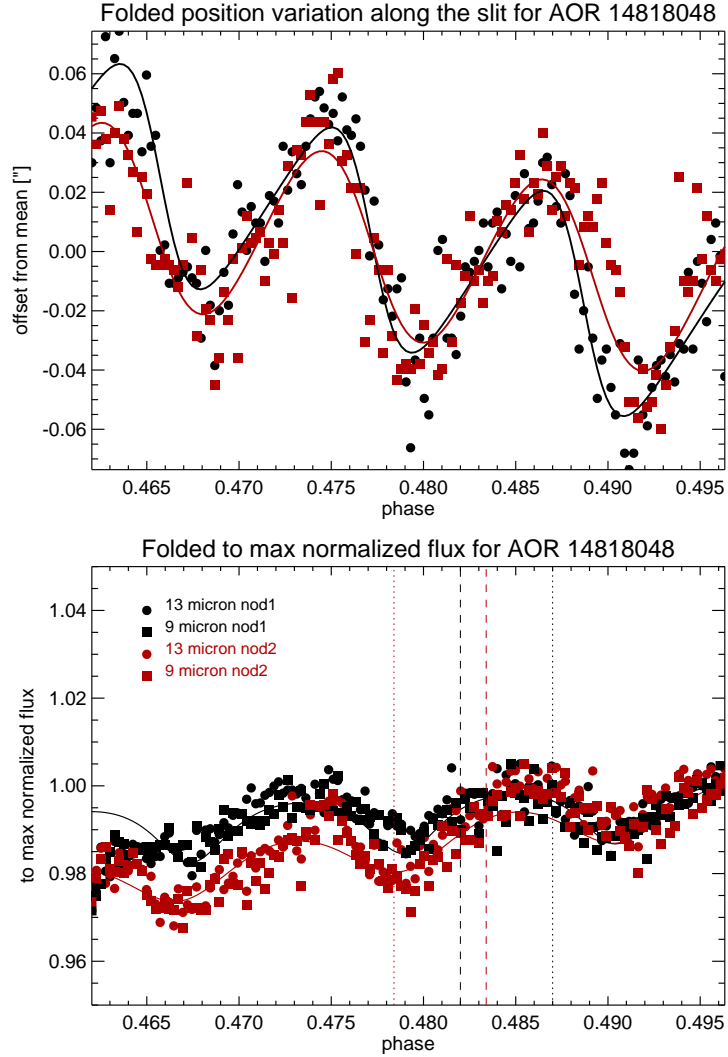}
\caption{{\bf Top:} Source position data in the slit spatial axis,
together with an elliptical pointing model for AOR 14818048.  Both
periodic motion and a linear drift are clearly identifiable in the
source position data.  The data are fit to determine the amplitude,
period, eccentricity, and linear drift terms.  After determining the
best fit pointing model using the data in the spatial axis, we
investigated the effect of the cross-slit motion predicted by this
model. {\bf Bottom:} The normalized cross-slit projected motion from
the elliptical pointing model as a function of time; with the
exception of the initial few points, the excellent agreement between
the model and the data is evident.  This strongly suggests that
time-variable pointing errors (periodic changes and linear drifts) are
primarily responsible for changing the measured flux.
\label{fig:positionFit}}
\end{center}
\end{figure}

\begin{figure}[h!]
\begin{center}
\epsscale{0.5}
\includegraphics[width=4in,angle=0]{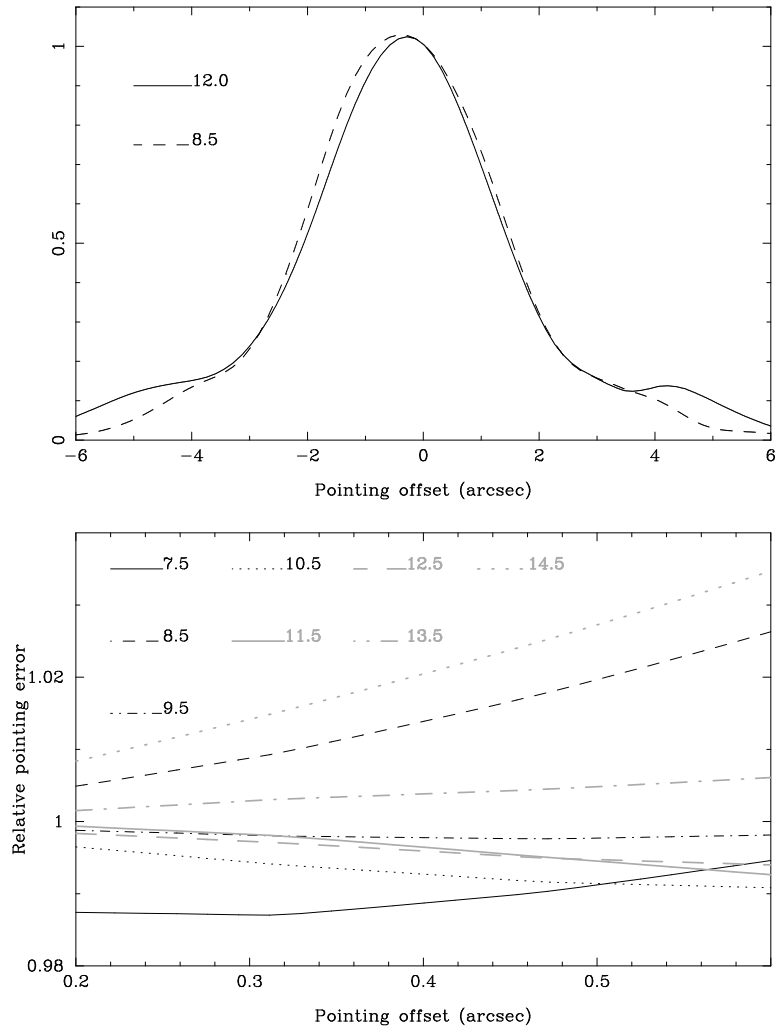}
\caption{{\bf Top:} The change in measured flux as a function of
source position in the cross-slit direction of the IRS entrance slit.
This calibration was developed from slit scans of IRS calibrators (see
text).  The measured flux does not peak at zero offset because of an
asymmetry in the wings of the PSF.  Given the 0.4 arcsecond radius of
the IRS ``high precision'' peakup mode error circle, the uncertainty
associated with the standard IRS high-precision peakup results in a
$\sim$ 7 \% uncertainty in the measured flux.  The measured flux error
due to the cross-slit pointing error is, to first order, common mode
in wavelength. {\bf Bottom:} The pointing offset correction as a
function of wavelength normalized to 12 $\mu$m.  This plot shows that
there is a spectral error of $\sim$ 1.5 \% across the instrument
passband due to the pointing offset effect for the standard
high-precision peakup.  The horizontal scale is set to include the
range of offsets present in the observations of HD 209458.
\label{fig:pointingCorrect}}
\end{center}
\end{figure}

While a periodic pointing error component is frequently seen in
Spitzer observations, it is not necessarily repeatable in terms of
shape or amplitude \citep{carey07}.  The IRS data we analyzed for HD
209458 show a periodic modulation of the measured flux (see Fig.
\ref{fig:initialTimeSeries}) that could be due to the Spitzer pointing
error.  To test the hypothesis that changes in the measured flux are
due to pointing errors, we modelled the pointing error periodic motion
in both the spatial and spectral axis.  This leads to an elliptical
motion that creates a symmetric profile about individual maxima and
minima.  The asymmetric profiles in these data require the addition of
a harmonic term for angular velocity; when this is incorporated, the
pointing error is given by
\begin{equation}
\dot{\theta} = \dot{\theta_{0}} + A_{\theta} sin(\omega t-\phi_{\theta}) ,
\end{equation}
\begin{equation}
x = x_{o} + m_{x} t + A_{x} cos[\omega \theta(t)-\phi_{x}] ,
\end{equation}
\begin{equation}
y = y_{o} + m_{y} t + A_{y} cos[\omega \theta (t)-\phi_{y}] ,
\end{equation}
where $t$ is time, $x$ is the position parallel to the slit axis (the
spatial dimension on the array), $y$ is the position perpendicular to
the slit axis (the spectral dimension of the array), $x_{o}$ and
$y_{o}$ are initial offsets, $m_{x}$ and $m_{y}$ are the linear drift
terms, $\dot{\theta}$ is the angular velocity, $A$ is the amplitude,
$\omega$ is the frequency, and $\phi$ is the phase.  The normalization
of $t$ and $\dot{\theta}$ is determined by the conditions
\begin{equation}
t \in [0,2\pi] ,
\int_0^{2\pi} \dot{\theta} dt = 2\pi ,
A_{\theta} \leq \dot{\theta_{0}} .
\end{equation}
\noindent We determined the parameters for the $x$ and $\theta$
components of our pointing model by fitting to the source motion along
the slit axis using the following steps for the data in each AOR:

\begin{enumerate}
\item{\bf Determine position:} For each measurement in the time
series, we constructed the spatial profile at each spectral channel.
These profiles are normalized by wavelength in the spatial axis and
shifted so that they can be ``stacked'' coherently.  A median spatial
profile is then determined.  This median spatial profile is then fit
with the function $sinc^{2}(x)$.  The fitted position of the maximum
of $sinc^{2}(x)$ as a function of time is used as the measure of
telescope pointing changes in the spatial axis.

\item {\bf Linear fit:} We fit and removed the linear component,
$m_{x}$, of the source position in the spatial axis of the data in
each nod.

\item{\bf Determine the frequency:} To determine the frequency,
$\omega$, of the periodic oscillation in each nod, we took the Fourier
transform of the linearly detrended position function in the slit
spatial axis.  The normalized frequency values were the same within
the errors, and the mean frequency was used in the remaining analysis.

\item{\bf Characteristic profile:} We folded the data, computed a
median profile and local standard deviation, applied a 10 $\sigma$
clip to remove discrepant points, and determined the mean profile.

\item {\bf Determine model parameters:} We determine the values for
$x_{o}, A_{\theta}$, $A_{x}$, $\phi_{\theta}$, $\phi_{x}$ by fitting
the predicted position along the slit axis to the measured source
position along the slit axis (see Fig. \ref{fig:positionFit}); the
values for these parameters are given in Table 1.
\end{enumerate}

\noindent At this point, we only need to determine the values of
$(y_{0},m_{y},A_{y},\phi_{y})$ to completely describe the telescope
pointing.  Judicious selection of values for
$(y_{0},m_{y},A_{y},\phi_{y})$ produces an estimate of the cross-slit
position changes that (with appropriate normalization) agree
remarkably well with the intensity time series (see Fig.
\ref{fig:positionFit} bottom) and successfully reproduce the
asymmetric component in the shape of the periodic modulation.

The excellent agreement between our simple pointing model and the
observed changes in the measured flux confirm that, in the case of a
point source, the IRS measurement of the flux density is affected by
the position of the (stellar) image in the spectrometer entrance slit.
The results (see Fig. \ref{fig:positionFit} bottom) imply that
pointing changes as small as $\sim$ 10 milliarcseconds have an effect
on the measured IRS flux for bright, point-like objects.  Because the
PSF size is a function of wavelength, the pointing error effect is
chromatic.  The asymmetry in the PSF wings also causes pointing errors
to be asymmetric with respect to the nominal center of the slit (this
can be seen in Fig. \ref{fig:pointingCorrect}).

Equipped with equations 1 and 4, we can now decompose the changes in
the measured flux density into three specific kinds of pointing
errors, all of which can be seen in Fig. \ref{fig:initialTimeSeries}.
Each of these pointing errors contributes a specific component of $y$.
Note that the values of $A_{y}$ \& $\phi_{y}$ are the same for all
AOR/nod combinations, while $y_{o}$ \& $m_{y}$ are different for each
AOR/nod combination.

\begin{itemize}
\item initial peakup/nod error - The pointing error associated with
the initial peakup or nod operation.  The high-accuracy peakup, used
for these observations, has a 1-$\sigma$ error circle radius of 0.4
arcseconds.  This translates into a flux uncertainty of $\sim5-10$\%.
When a nod is executed, there is significant motion perpendicular to
the slit axis.  This is the reason why the median flux density
differs in each AOR/nod1.  The initial error is static and represents
a constant offset described by $y=f(y_{0})$.
\item pointing drift - During an observation, there is a slow linear
drift in pointing during each nod.  The drift rate is larger at the
nod2 position.  The slow pointing drift rate ranges from 3 mas/hr to
19 mas/hr (based on a 1.85 arcsecond per pixel plate scale and a nod
duration of 2.9026 hr - see Tab. 1). This linear pointing error is
described by $y=f(m_{y})$.
\item periodic error - The Spitzer telescope has a known periodic
pointing error $\sim \pm 30$ milliarcseconds.  This is the error that
causes the clear periodic modulation of the flux.  The periodic
position changes are described by $y=f(A_{y},\phi_{y})$.
\end{itemize}

\subsubsection{Background Correction} 
In the mid-infrared, accurate measurement of the infrared source flux
requires subtraction of the background due to local zodiacal emission.
To remove the background contribution to the spectrum, we construct
and subtract a median background image.  However, this median image
must be constructed with care as there is a systematic error in the
backgound estimate due to leakage from the bright source.  This
leakage is manifested as a flux density offset between the background
at the nod1 and nod2 positions.  In principle, this offset could be
caused by structure in the background.  However, inspection of IRS
calibrator star data shows that the difference in the background
between the nods is systematic in that it occurs for all the multiply
observed IRS calibrators we checked; the effect is highly repeatable
and is proportional to the measured source flux.  IRS calibrators
observed with a series of slit offsets show the measured source flux
decreases with the slit offset from the target, and the background
offset is proportional to the measured flux.  This suggests that some
of the light from the source contaminates the background through the
wings of the PSF.  Because the Spitzer PSF is asymmetric
\citep{bayard04}, the leakage differs in nod 1 and nod 2.

To determine the amount of a point source contamination in the
background estimate, we have used observations covering an interval of
approximately three years for five IRS calibrator stars (HR 6606, HR
7341, HD 166780, HD 173511, HR 6348), together with the assumption of
a locally uniform background.  The IRS calibrators we selected were
observed in the nominal nod1 and nod2 positions for both SL1 and SL2
modes.  These stars were observed on a regular basis throughout the
Spitzer operational period up to the time of these observations.  Each
star was typically observed at least 20 times over a three year
interval.  Thus, slit precession over a period of one year is a strong
test of our assumption of uniform background.

\begin{figure}[h!]
\begin{center}
\epsscale{0.5}
\includegraphics[width=4in,angle=0]{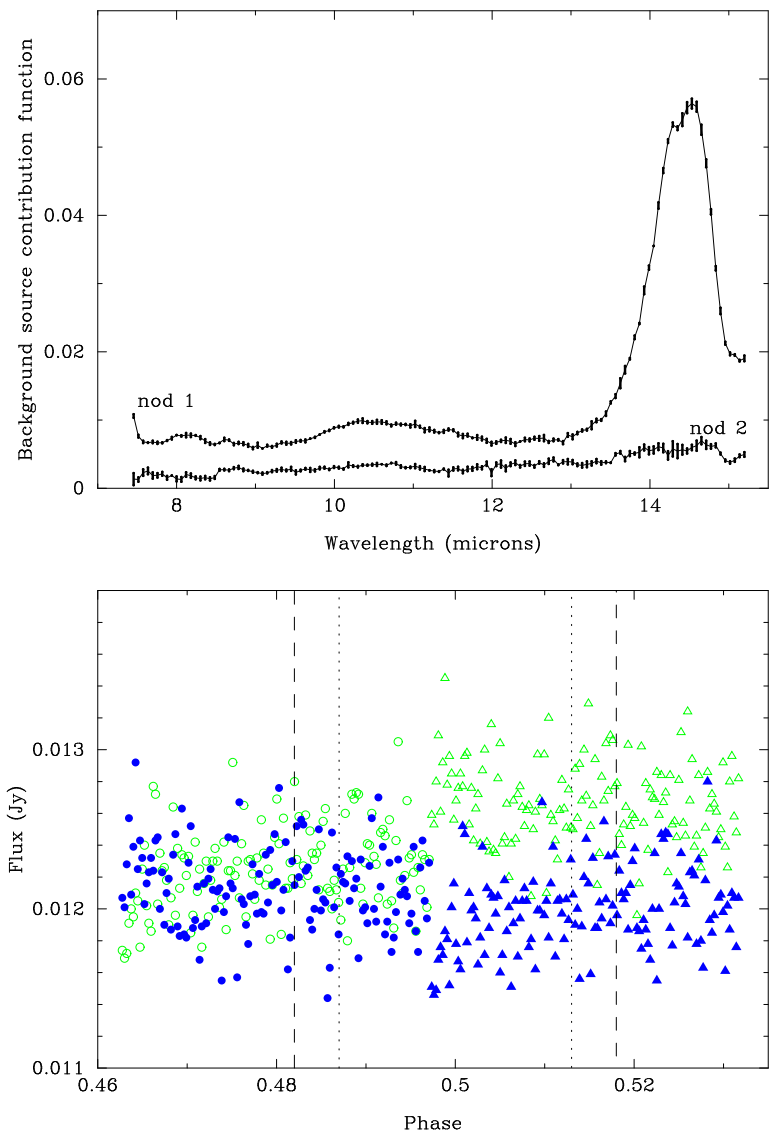}
\caption{{\bf Top:} The background source contribution function (BSCF) is
different for the two nods because of an asymmetry in the telescope
PSF (see text for description of the derivation).  Multiple stars and
epochs were used to determine both the correction and the
uncertainty. {\bf Bottom:} The open circles (nod1 background) and
triangles (nod2 background) are the estimated background prior to
correction for the source contribution at 12 $\mu$m in AOR 14818048.
The filled circles and triangles show the result of applying the
correction for source leakage.
\label{fig:background}}
\end{center}
\end{figure}

We determined the source contamination in the background by
subtracting two SL1 background positions when the star is in the SL1
and SL2 positions.  The background source contribution function,
$BSCF$, for the nod1 position has the form
\begin{equation}
BSCF_{nod1} = \frac{S_{leak}(nod1)}{S_{source}(nod2)} = 
\frac{B_{nod1}(SL1,nod2) - B_{nod1}(SL2,nod1)}{S'(nod2) -
B_{nod2}(SL2,nod1)} \left( \frac{RSRF_{nod2}}{RSRF_{nod1}} \right),
\end{equation}
where $B_{nod1}(SL1,nod2)$ is the background at the nod1 position
measured when the source is located at the SL1, nod2 position;
$B_{nod1}(SL2,nod1)$ is the background at the nod1 position measured
when the source is located at the SL2, nod2 position; $S'(nod2)$ is
the measured source flux at the SL1, nod2 position with no background
correction; and $RSRF$ is the relative spectral response function at
either the nod1 or nod2 source position.  For each term, the subscript
denotes the position on the array where the value was measured while
the source position at the time of the measurement is indicated by the
parenthesis.  Thus, $B_{nod1}(SL2,nod1)$ is the background measured at
the SL1, nod1 position when the source is located at the SL2, nod1
position.  Since we are calibrating SL1 data, the background is always
measured in the SL1 slit.  However, determining the $BSCF$ requires
using data when a star was observed with both the SL1 and SL2 slits.
The $BSCF_{nod2}$ is similarly defined except that all nod1 instances
become nod2 and {\it visa versa}.  The corrected background
flux density at the two SL1 nod positions is then
\begin{equation}
S(nod1) = \left[ S'(nod1) - B_{nod1}(SL1,nod2) \right] + BSCF_{nod1} \times S(nod2),
\end{equation}
and
\begin{equation}
S(nod2) = \left[ S'(nod2) - B_{nod2}(SL1,nod1) \right] + BSCF_{nod2} \times S(nod1).
\end{equation}
The system of linear equations is then solved for the background
corrected, measured source flux density, $S$, at each nod position.
This results in an estimate of the $BSCF$ each time the calibrator
stars were observed.  We then averaged the results for all the
calibrator observations to determine a mean $BSCF$ (see
Fig. \ref{fig:background}).  The uncertainty in the $BSCF$ at each
wavelength was determined by the standard deviation in the mean.

Applying the $BSCF$ substantially reduces the background flux density
offset between the nod1 and nod2 positions (see
Fig. \ref{fig:background}).  For wavelengths shorter than 13 $\mu$m,
the correction we derive is $\sim$0.4\% for nod 1 and $\sim$0.9\% for
nod 2.  This means that there can be $\sim$ 0.5\% of the source flux
present in the wings of the PSF $\sim$ 20 arcseconds away from the
observed source position.  Thus, the contamination of the background by
the source is of the same magnitude as the signal from the exoplanet.
This correction for source contamination of the background may not be
necessary for many observations.  However, for high dynamic range
measurements on bright point sources, neglecting the leakage of the
source into the background estimate introduces systematic errors in
the data for each nod.

\subsubsection{Latent Charge Accumulation} 
In the context of the IRS instrument, latent charge accumulation has
been reported by several authors
\citep{grillmair07,richardson07,deming06}, and is sometimes termed
``charge trapping''.  Currently, the details of the semi-conductor
physics that produce the effect are not well understood.  Empirically,
the responsivity of a pixel initially depends on the illumination.
When the flux density time series is median-normalised (e.g.
$F_{i}(\lambda)/<F(\lambda)>$), the light curves at each wavelength can
be ``stacked'' \citep{richardson07}.  The effect of latent charge
accumulation can be seen at the beginning of each AOR in
Fig. \ref{fig:initialTimeSeries}; the effect is characterized by a
rapid initial increase in the measured flux density, which then
approaches an equilibrium.  If one excludes the first $\sim $ 20
points in each AOR and finds the slope of a best fit line to the data,
the slope in nod2 is greater than the slope in nod1.  As
Fig. \ref{fig:positionFit} shows, a simple pointing model explains the
changes in the measured flux after the first $\sim$ 20 minutes.  Note
that it is possible to confuse the linear component of the pointing
drift with the effect of latent charge accumulation after the first
$\sim$ 20 minutes.  By explicitly modelling the pointing, our analysis
breaks this degeneracy and allows us to separate the effects of these
two systematic errors.  We conclude that latent charge effects are
negligible after the first 20 minutes.  We omitted the data affected
by latent charge accumulation from further analysis so that the
effects of latent charge do not impact our estimate of the planet
spectrum.

\subsection{Spectral Response Function}

After the initial extraction, the data were background corrected using
the background correction discussed above.  The next stage of the
calibration was to derive and apply a spectral response function.
Using the IRS calibrator $\eta$ Dor, we derived our own spectral
response function, the $SRF$, for the nominal nod positions and
extraction aperture.  We selected this source for defining the
spectral response function because it has the same brightness as HD
209458 and thus should minimize any remaining instrument residuals.
Both HD 209458 and the $\eta$ Dor data were extracted using identical
methods, and both incorporate identical methods for background
correction.  Thus, the treatment of both the calibrator and source data
sets is fully self-consistent.  In the case of AOR 14818048, an
additional calibration step for the $SRF$ was required because the
observations were not carried out at the nominal nod positions.  To
determine the changes in the $SRF$ for other (but still relatively
nearby slit positions), we used observations of the IRS calibrator
stars HD 42525 and HR 7341, which were observed at intervals along the
IRS slit.  We interpolated between observing positions to determine
how the calibrator star's spectrum changed as a function of slit
position and used this information to renormalize the $SRF$ derived,
using $\eta$ Dor for the nod positions used in AOR 14818048.

At this point, the data were ready for extraction of the spectrum.  Of
the three major systematic errors, the background had been removed (at
this point in the calibration sequence) by explicit calibration.  The
effects of latent charge were removed by excluding the effected data
from the spectral estimation.  However, the pointing error remained
uncorrected.  In what follows, two methods were used to correct the
pointing error and extract the planet spectrum.

\subsection{Spectral Estimation}

\subsubsection{Differential Method} 
This approach assumes that changes in the measured flux have a
wavelength-independent component, characterized by $G(t)$, which can
vary on a timescale of minutes, and a wavelength-dependent component,
$G(\lambda)$, which is stable for a given nod but can change between
nods.  The $G(\lambda)$ term is removed by construction of a spectral
flat.  We derived a spectral flat for each nod by comparing the
average flux in each spectral channel to the flux, $F(\lambda)$, of a
stellar photosphere model for HD 209458 \citep{kurucz92} normalized to
the 12 $\mu$m flux.  Thus at each wavelength, $\lambda$, the spectral
flat was defined as the inverse of
$[S_{S}(\lambda)/F_{Kurucz}(\lambda)]\times[F_{Kurucz}(12)/S_{S}(12)]$
where $S_{S}$ is the measured flux observed in interval S.  After
normalization of each nod by the associated spectral flat field, the
data were assumed to vary only in time; a more extensive discussion of
this technique can be found in \cite{bryden06} and \cite{beichman06}.
To reject the wavelength-independent $G(t)$ term, we constructed a
differential observable using the following method.  During period SP
(star+planet), the measured flux, $S_{SP}(\lambda)$, can be written as

\begin{equation}
S_{SP}(\lambda) = G(t) \left[ F_{S}(\lambda) + F_{P}(\lambda)\right].
\end{equation}

\noindent This can be expanded as

\begin{equation}
S_{SP}(\lambda) = G(t) F_{S}(\lambda) \left[ 1 +
\frac{F_{P}(\lambda)}{F_{P}(\lambda^{'})} 
\frac{F_{P}(\lambda^{'})}{F_{S}(\lambda^{'})}
\frac{F_{S}(\lambda^{'})}{F_{S}(\lambda)} \right],
\end{equation}

\noindent where $F(\lambda)$ is the true source flux, the
subscripts refer to the star or planet, and $\lambda^{'}$ is a
reference wavelength selected for the comparison.  We set the transit
depth at $\lambda^{'}$ to a plausible value such that
$\beta=(F_{P}(\lambda^{'})/F_{S}(\lambda^{'}))<<1$ and the transit
depth at $\lambda$, relative to the transit depth at $\lambda^{'}$, is
$\alpha=(F_{P}(\lambda)/F_{P}(\lambda^{'}))$. $S_{SP}(\lambda)$ can
then be expressed in terms of $\alpha$ and $\beta$ as

\begin{equation}
S_{SP}(\lambda) =G(t) F_{S}(\lambda) \left[ 1 + \alpha \beta
\frac{F_{S}(\lambda^{'})}{F_{S}(\lambda)} \right].
\end{equation}

\noindent During period S (star only), the measured signal, $S_{S}$, is
$S_{S}(\lambda) = G(t) F_{S}(\lambda)$. The ratio of the two
wavelengths during the SP and S periods is
$R_{SP}=S_{SP}(\lambda)/S_{SP}(\lambda^{'})$ and
$R_{S}=S_{S}(\lambda)/S_{S}(\lambda^{'})$.  The advantage of taking
the ratio is that the wavelength-independent gain term, $G(t)$, drops out. 
Appropriate substitution, and solving for $\alpha$, yields

\begin{equation}
\alpha = [R_{SP} (1 + \beta) - R_{S}] / \beta .
\end{equation}

\noindent The observables are $R_{SP}$ and $R_{S}$, and $\alpha$ is
the measure of the brightness of the planet at $\lambda$ compared to
the planet brightness at $\lambda^{'}$.  The results in
Fig. \ref{fig:specCompare} reflect a value for $\beta=0.003$.
However, the results for the spectral slope are not strongly dependent
on the assumed value for $\beta$, and we explicitly measured the
eclipse depth in any case (using the absolute method).

We summarize the steps in the differential method as follows:
\begin{enumerate}
\item Treat each AOR and nod combination as an independent secondary
eclipse measurement with an independent calibration; this leads to
four independent estimates of the planet spectrum.
\item Normalize the data in a given nod by dividing the flux
density by the 12 $\mu$m value at each sample in the time series.
\item Average the SP and S intervals in the time series to get the
``star only'' and ``star+planet'' spectra.
\item Normalize a Kurucz model for HD 209458 flux density by the model's
12 $\mu$m prediction.
\item Construct a ``super flat'' by dividing the normalized Kurucz
model into the normalized data (for both SP and S intervals).
\item Estimate the planet spectrum by subtracting the S interval 
spectrum from the SP interval spectrum.
\end{enumerate}
The four estimates of the exoplanet spectrum are then averaged to
create the final spectrum.  The errors are estimated by taking the
average value of the differences in the estimate at each wavelength.

To assess the magnitude of residual systematics, we made a comparison
(see Fig. \ref{fig:specCompare}) between the spectra from each of the
four independent AOR and nod combinations.  In the central region of
the IRS SL1 instrument bandpass, the spectra are in relatively good
agreement.  However, this agreement becomes worse at either end of the
instrument bandpass; this is especially true for wavelengths between
7.5 and 9 $\mu$m.  The reason for this is that the assumption that
$G(\lambda)$ is time invariant is only a first order approximation.
Because we have normalized by the 12 $\mu$m flux density, the effect
of the small, uncorrected pointing errors within a nod is greatest at
the band edges (see Fig. \ref{fig:pointingCorrect}).  To determine the
best estimate of the differential spectrum, the four independent
differential spectra are averaged together.

\subsubsection{Absolute Method:} Here we describe how to apply the $\zeta$
correction to the source and calibrator data.  Although we do not
know {\it a priori} what the telescope pointing error is, we can
determine the correct flux density for a given pointing error, $y$,
using $F=S/\zeta(y)$.  From Eq. 4, we know that $y =
f(y_{0},m_{y},A_{y},\phi_{y})$.  Thus, our task is to identify the
correct values for ${y_{0},m_{y},A_{y},\phi_{y}}$.  One way to do this
is to require that the absolute spectrum be self-similar.  We
implemented this by constructing all unique combinations of the
relation
\begin{equation} 
{\bf R(i,j)} =
F_{i}(\lambda,\theta)/F_{j}(\lambda,\theta) =
\frac{S(t_{i}) \zeta(y_{j})}{S(t_{j}) \zeta(y_{i})}
\end{equation}
for the SP and S portions of each nod separately, where $i$ and $j$
are individual measurements in the time series.  We then iteratively
searched this space to determine the values of
${y_{0},m_{y},A_{y},\phi_{y}}$ (given in Tab. 2),
which resulted in most closely approximating ${\bf
R(i,j)}=1$. Fig. \ref{fig:timeSeries} shows the result of the
application of the pointing offset correction, and the secondary
eclipse event is directly visible.  Similarly, we applied the $\zeta$
correction to $\eta$ Dor and determinded the pointing offset by
requiring spectral self-similarity.  As with the differential method,
we evaluated the internal consistency of the pointing correction by
comparing the spectra from both nods in both AORs.  The agreement
between the absolute spectra is excellent (see
Fig. \ref{fig:method2check}), and we now compare the
differential and absolute spectra to assess the level of any residual
systematics.

\section{Discussion}

In this section, we compare the results of the differential and
absolute methods we used to extract the planet spectrum.  The
assumption of wavelength-dependent stability used in the differential
method is evaluated.  We discuss our estimate of the eclipse
depth and spectral features; we interpret these results in the context
of recent models.  We also discuss the significant differences between
our analysis methods and results and those of \cite{richardson07}.

\subsection{Comparison of differential and absolute methods}

A fundamental strength of our approach is the use of two
semi-independent methods to demonstrate understanding and calibration
of the dominant systematic errors.  As Fig. \ref{fig:specCompare}
shows, the agreement between the differential and absolute planet
spectrum estimation methods is excellent over most of the instrument
passband.  While agreeing within the errors, between 7.5 and 9 $\mu$m,
the differential spectrum is systematically below the absolute spectrum.
This is caused by small pointing errors occuring within a nod that are
not removed by the differential method.  Because the internal scatter
of the absolute method is similar at all wavelengths, we consider the
absolute spectra to be the best estimte of the planet spectrum.

That the two spectral extraction procedures yield consistent results
is encouraging and gives us a high degree of confidence that the
calibration of systematic errors is successful within the error
bars. Given that the differential method appears to make no specific
correction for pointing, one might wonder why the agreement with the
absolute method is so good.  The source of the agreement is that the
normalization by the Kurucz model corrects for any chromatic error,
pointing or otherwise, {\it so long as the chromatic error changes in
time are relatively small}.  Thus, normalization by the Kurucz model
corrects the chromatic error produced by the largest pointing errors,
which are static and occur during the initial peak-up and during the
nod.  Because the periodic pointing errors are relatively small, the
change in the measured flux during a nod is, to first order,
wavelength independent, and thus the spectral flat field is a good
approximation for the flux correction due to the initial pointing
error.  Thus, the agreement between the differential and absolute
spectral estimation methods supports the original assumption that the
$G(\lambda)$ term is relatively (but not completely) constant during a
nod.  The increased size of the error bars in the differential method
results from the periodic component of the pointing errors.

\subsection{Eclipse Depth}

We have determined an average eclipse depth for the data by
normalizing the absolute (pointing error corrected) flux density time
series at each wavelength by the median values of the time series,
$F(\lambda)/<F(\lambda)>$.  This is then averaged over wavelength to
develop a broad-band light curve.  The result of this can be seen in
Fig. \ref{fig:eclipseDepth}; the broad-band light curve clearly shows
the eclipse and the transitions between ingress and egress.  We can
derive four independent estimates (one for each nod) of the broad-band
eclipse depth, and these are consistent within the errors.  After
averaging the individual estimates, we find the average eclipse depth
between 7.6 and 14 $\mu$m to be 0.00315$\pm$0.000315.  This minor
restriction in wavelength was implemented to exclude the channels with
lower SNR.  When compared to theoretical models \citep{burrows06}, the
measured eclipse depth suggests that substantial heat redistribution
from the dayside to the nightside is occurring.  This evidence of heat
redistribution is similar to the interpretation given to observations
of HD 189733b by \cite{grillmair07} and \cite{knutson07}.

\subsection{Planet Spectrum}

We have determined the planet spectrum (see
Fig. \ref{fig:absoluteSpec}) and find it to range from about 600
$\mu$Jy at 7.5 $\mu$m to about 200 $\mu$Jy at 15.2 $\mu$.  The SNR in
the spectrum ranges from $\sim$ 10 at short wavelengths to $\sim$ 2 at
the longest wavelengths.  To our knowledge, this is the first
determination of the absolute spectrum of exoplanet emission.  Our
results for the spectral shape agree well with previous work (see
Fig. \ref{fig:richardsonSpec}) albeit with improved SNR; specifically,
we confirm the marginal detection by \cite{richardson07} of a narrow
feature near 7.7 $\mu$m.  For most wavelengths, the planet spectrum is
characterized by approximately featureless emission.  However, between
7.5 and 8.5 $\mu$m, there is evidence for one broad (previously
unreported) and one narrow (previously reported) spectral feature.

Both the absolute spectrum and the contrast spectrum show evidence of
a possible $\sim$ 0.5 $\mu$m-wide feature centered around 8.1 $\mu$m,
with a significance of about 4 $\sigma$.  This broad feature
represents a flux deficit from the local trend and could be due to
absorption.  At the full spectral resolution, there is also a
suggestion of a narrow feature around 7.7 $\mu$m.  This narrow feature
candidate could be either in absorption (a deficit relative to the
local trend in the 7.57 and 7.63 $\mu$m channels) or in emission (an
excess relative to the local trend in the 7.69 and 7.75 $\mu$m
channels).  The shape of the broad feature causes us to favor the
hypothesis of a narrow absorption feature in the 7.57 and 7.63 $\mu$m
channels.  However, movement of any one of these four spectral points
(7.57, 7.63, 7.69, and 7.75 $\mu$m) by $\sim$1.5 $\sigma$ towards the
local trend would convert this candidate feature into an outlier
consistent with a normal measurement distribution.  The narrow feature
candidate is sufficiently marginal that additional observations are
required to confirm or rule out a spectral feature at this wavelength.

The indication that the spectrum of HD 209458b contains one broad and
one narrow feature between 7.5 and 8.5 $\mu$m is supported by the
\cite{richardson07} measured spectrum.  Indeed, the striking
qualitative agreement (one broad and one narrow feature) between
previous work and our results for the spectral modulation between 7.5
and 8.5 $\mu$m is a strong indication that this modulation is real.
Although \cite{richardson07} did not discuss the broad feature, it is
present in their spectrum and we confirm their measurement.  While we
cannot totally exclude the possibility of some residual instrument
systematic, it is highly significant that the shape of this spectral
modulation is consistent using three independent methods conducted by
two independent groups.  Because of the repeatability of the result
and the maturity of the exoplanet spectrum determination, the spectral
modulation between 7.5 and 8.5 $\mu$m is likely real and may serve as
a useful constraint on models for emission from HD 209458b.

In the interpretation of the previous results for these data,
\cite{richardson07} reported the detection of a broad emission feature
centered at 9.65 $\mu$m, identified as a silicate feature, and a
narrow emission feature centered at 7.78 $\mu$m.  We find no evidence
to support the identification of a 9.65 $\mu$m feature in our
spectrum.  Additional averaging and scrolling median filtering does
not reveal any candidate feature with the characteristics claimed by
\cite{richardson07}.  It is possible that the narrow feature
identified by \cite{richardson07} corresponds to the 7.67 and 7.75
$\mu$m channels in our analysis.  If this is the case, the difference
in wavelength is possibly due to the non-standard wavelength
calibration method used by \cite{richardson07} (see discussion below).
However, we stress that the candidate {\it absorption} feature at 7.57
$\mu$m is at least as likely as an {\it emission} feature at 7.69
$\mu$m

\subsection{Differences with Previous Work}

There are several significant differences in our data calibration
method and results when compared to the approach used by
\cite{richardson07}.  Our approach explicitly corrects for the
telescope pointing error and source leakage into the background; both
of these effects are chromatic errors capable of introducting
systematic errors in a spectrum.  We also use two methods, one
differential and one absolute, to extract the spectrum of the
exoplanet, and we demonstrate good agreement between the methods.
Unlike the previous work, we are able to explicitly measure the
secondary eclipse depth from the IRS data.  The improved SNR and lower
internal scatter in our spectrum allows a clear identification of the
spectral modulation between 7.5 and 8.5 $\mu$m, and rules
out the possibility of significant silicate emission at 9.65 $\mu$m.
Below, we explain some of the important details in the differences
between our methods and results and those of \cite{richardson07}.

\begin{itemize}

\item {\bf absolute spectrum} (result): We have determined the
spectrum of HD 209458b in Jy.  To our knowledge, this is the first
absolute determination of an exoplanet emission spectrum.

\item {\bf eclipse depth} (result): We explicitly determine the
broad-band eclipse depth from the IRS data at high SNR ($10\sigma$).
This determines the eclipse depth in the IRS SL1 instrument passband
and avoids the uncertainty associated with incomplete matching of the
IRS wavelength coverage to the 8 $\mu$m IRAC channel.

\item{\bf spectral features} (result): We find no evidence for the
silicate feature identified by \cite{richardson07}.  There is the
possibility of a narrow candidate feature at $\sim$7.7 $\mu$m, but at
the 1.5-$\sigma$ level it is consistent with noise.  In addition, the
position of the 7.64 and 7.70 $\mu$m spectral points relative to the
neighbors make this candidate feature as likely to be an absorption
feature as an emission feature.  

\item {\bf wavelength calibration} (method): As part of the spectral
extraction process, using SMART, we include the wavesamp.tbl table
calibration file provided by the Spitzer Science Center.  This
approach implements an interpolation method to determine how
fractions of a pixel contribute to a given wavelength.  This approach
accounts for the spectra tilt and curvature and provides
Nyquist sampling of the spectra in the dispersion direction.  In
contrast, the wavelength definition used by \cite{richardson07} is
based on the b0\underline{ }wavesamp\underline{ }wave.fits file
which, according to the IRS handbook, is for notional purposes only
and should not be used for a scientific analysis.  It is likely that
relying on the b0\underline{ }wavesamp\underline{ }wave.fits file
for the wavelength definition is why the wavelength scales for the two
AORs are different in the \cite{richardson07} analysis.

\item {\bf background correction} (method): Our background subtraction
approach includes a correction for contamination from the source.
This is a wavelength-dependent effect, which is of the order of the
secondary eclipse depth.  Failure to correct for source leakage in a
normal background subtraction approach causes a wavelength-dependent
error if the data in a nod are simply adjusted (the ``multiplicative
factor'' for \cite{richardson07}) to make the time series continuous.

\item {\bf pointing correction} (method): Our method includes a
specific correction for the pointing error, which corrects the static
offset, periodic changes, and linear drift error terms in the telescope
pointing.  Uncorrected pointing errors that change with time
introduce spectral errors.

\item {\bf spectral response function} (method): Our determination of
the spectral response function includes a correction for both the
pointing error and the source contamination of the background.  The
spectral response function derivation is required for an absolute
exoplanet spectrum.

\item {\bf error estimate} (method): Our error bars are determined by
the standard deviation in the mean of multiply determined quantities
(e.g. the background corrected and pointing corrected time series) and
by the root sum of squares for combined quantities.  The error bars in the
\cite{richardson07} analysis are determined by offsetting the time
series by one time step, subtracting the original time series, and then
determining the standard deviation in the mean of the resulting time
series (in every spectral channel).  This approach removes the effect of
all systematic error with timescales longer than $\sim$ 2 minutes and
thus has the potential to underestimate the measurement uncertainty.

\end{itemize}

\section{Conclusions}

Our results for the spectrum of HD 209458b are consistent with a
smooth, largely featureless spectrum ranging from about 600 $\mu$Jy at
7.5 $\mu$m to about 200 $\mu$Jy at 14 $\mu$m.  However, there is
evidence of a spectral feature between 7.5 and 8.5 $\mu$m.  We find
evidence for a broad $\sim$ 0.5 $\mu$m wide feature, centered at
approximately 8.1 $\mu$m, that is possibly due to absorption.  Near 7.7
$\mu$m we find a narrow feature candidate that could be either
absorption or emission, depending on wavelength and local baseline trend
assumptions; this candidate feature is only $\sim$ 1.5 $\sigma$ from
being consistent with noise.  We find no evidence for the silicate
feature reported in \cite{richardson07}. The relatively
smooth character of the HD 209458b spectrum suggests the planet
emission is dominated by purely thermal emission over most of the IRS
SL1 passband.  However, the spectral modulation between 7.4 and 8.4
$\mu$m is significant and suggests that the dayside vertical
temperature profile of the planet atmosphere is not entirely
isothermal \citep{fortney06}.

We are able to make a direct measurement of the eclipse depth.
Between 7.6 and 14.2 $\mu$m we find an average eclipse depth of
0.00315$\pm$0.000315; when compared to planet emission models such as
\cite{burrows07}, the measured eclipse depth is suggestive of
substantial heat redistribution between the nightside and dayside.
Similar conclusions have been drawn for observations of HD 189733b
\citep{grillmair07,knutson07}.

The methods we have developed for calibration of the background and
pointing errors represent a significant improvement in the
state of the art for IRS calibrations on bright objects.  Using a
simple pointing model and requiring self-consistency of the spectrum
for the ``star+planet'' and ``star'' portions of the time series, we
are able to optimally recover the spectrum of HD 209458b.  By applying
our calibration of ({\it i}) source contribution to the background and
({\it ii}) pointing errors to the definition of the spectral response
function, we have achieved an absolute flux density calibration
approaching 0.1 $\%$.  This implies that our calibration method is
suitable for spectroscopy of emission from the nightside of exoplanets
and would significantly increase the SNR for IRS spectra of relatively
bright point sources.

\acknowledgments

We thank the original PI team for the proposal and preparation of the
AORs required to obtain these data.  We appreciate the comments of the
anonymous referee who encouraged us to fully describe our calibration
process and extend our calibration method to the entire IRS SL1
passband; these suggestions led to the detection of spectral
modulation that was outside our original spectral passband.  We also
thank Drake Deming for several helpful conversations regarding the
reduction of secondary eclipse data.  We thank John Bayard for several
helpful discussions concerning Spitzer pointing errors and Sara Seager
for discussion regarding the possible role of clouds in exoplanet
atmospheres.

\begin{figure}[h!]
\begin{center}
\epsscale{0.9}
\includegraphics[angle=-90,scale=0.7]{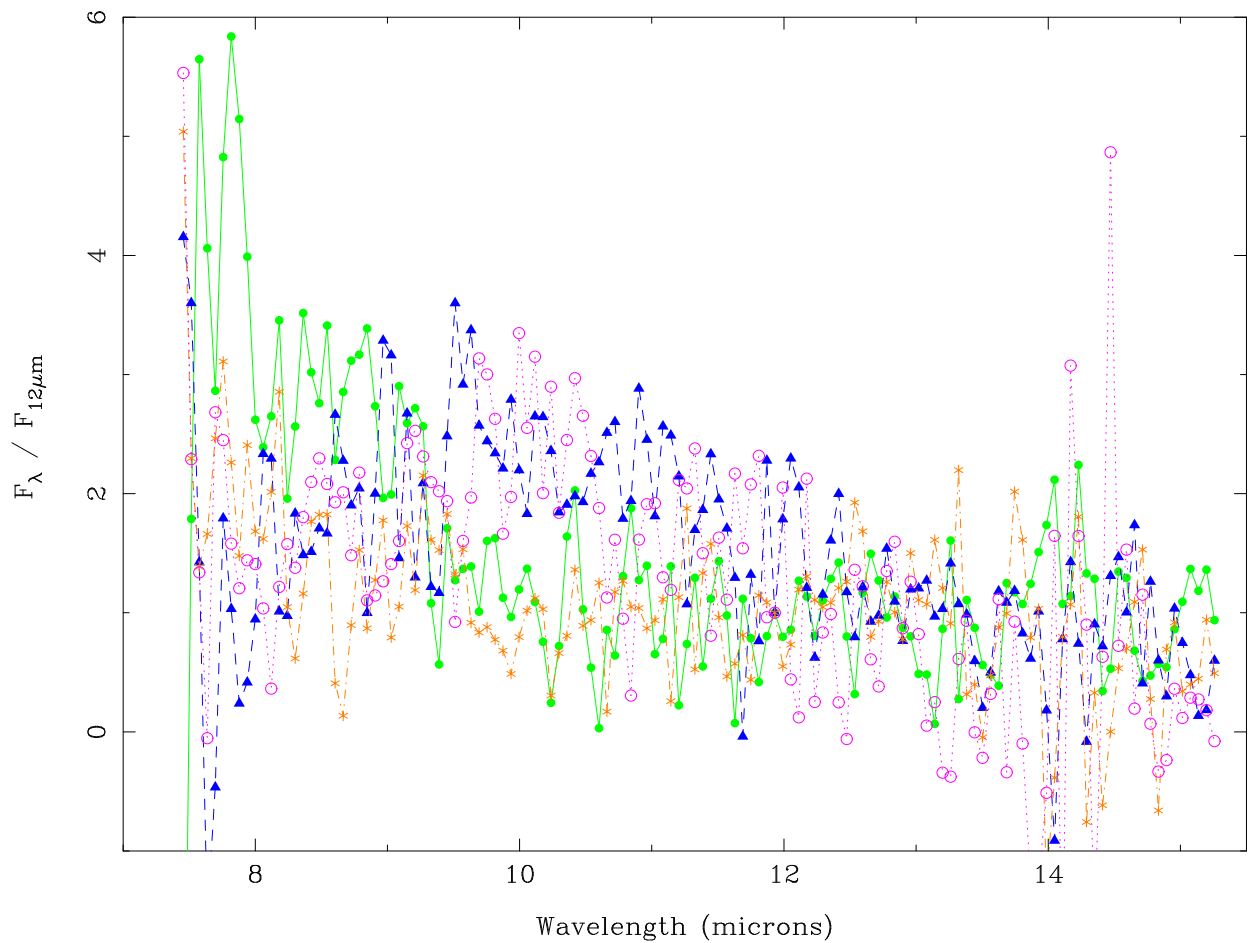}
\caption{To determine the level of internal consistency for the
differential method, we have plotted the four independent estimates of
the planet's spectrum.  The internal agreement is good over most of
the instrument bandpass.  However, this internal agreement is
substantially worse between 7.5 and 9 $\mu$m; this is due to small,
uncorrected pointing effects that are greatest at the edges of the
passband because we have normalized by the 12 $\mu$m flux density.
\label{fig:method1check}}
\end{center}
\end{figure}

\begin{figure}[h!]
\begin{center}
\epsscale{0.5}
\includegraphics[width=4in,angle=-90]{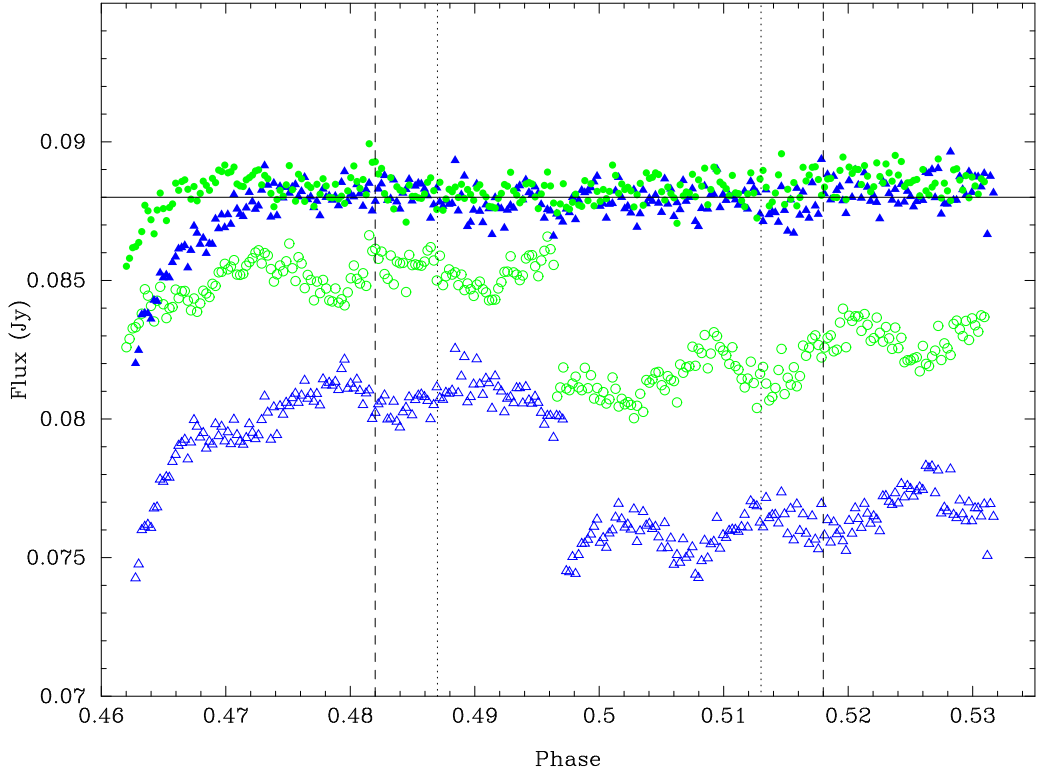}
\caption{The 12 $\mu$m flux time series for AOR 14817792 and 14818048
as a function of orbital phase.  The background-subtracted data are
shown in open circles and triangles.  The large discontinuity halfway
through each data set corresponds to a nod; periodic changes and
drifts also affect the time series measurements.  The changes in
measured flux are primarily due to telescope pointing errors, which
cause variable vignetting by the spectrometer entrance slit of the
wings of the source point spread function.  The filled points show the
pointing-corrected flux.  The signature of the secondary eclipse is
visible as an interval with systematically lower flux density.  The
vertical dashed lines indicate the region of ingress/egress where the
planet is partially obscured by the star.  A horizontal line has been
added to aid in identifying the flux density decrease during secondary
eclipse.  Note that the effect of latent charge accumulation can be
seen in the apparently increasing flux in the initial $\sim$ 20
points for each AOR (near orbital phase 0.464).
\label{fig:timeSeries}}
\end{center}
\end{figure}

\begin{figure}[h!]
\begin{center}
\epsscale{0.9} 
\includegraphics[angle=-90,scale=0.7]{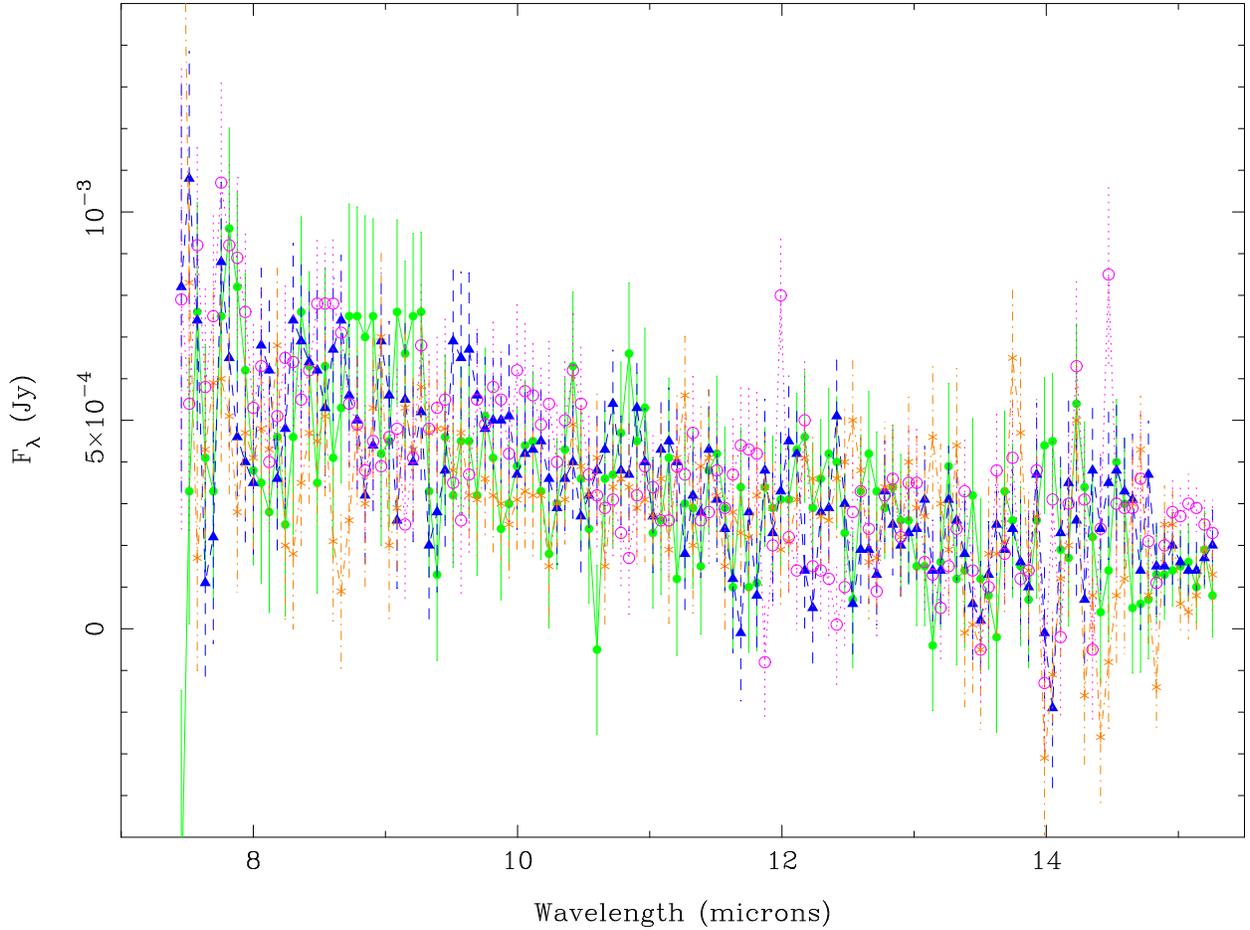}
\caption{Four independent estimates (nod1 and nod2 for both AOR
  14817792 and AOR 14818048) of the planet spectrum, determined using
  the absolute method, are plotted together as an internal consistency
  check.  The internal agreement of the absolute method is excellent
  over the entire instrument passband.
\label{fig:method2check}}
\end{center}
\end{figure}

\begin{figure}[h!]
\begin{center}
\epsscale{0.9} 
\includegraphics[angle=-90,scale=0.7]{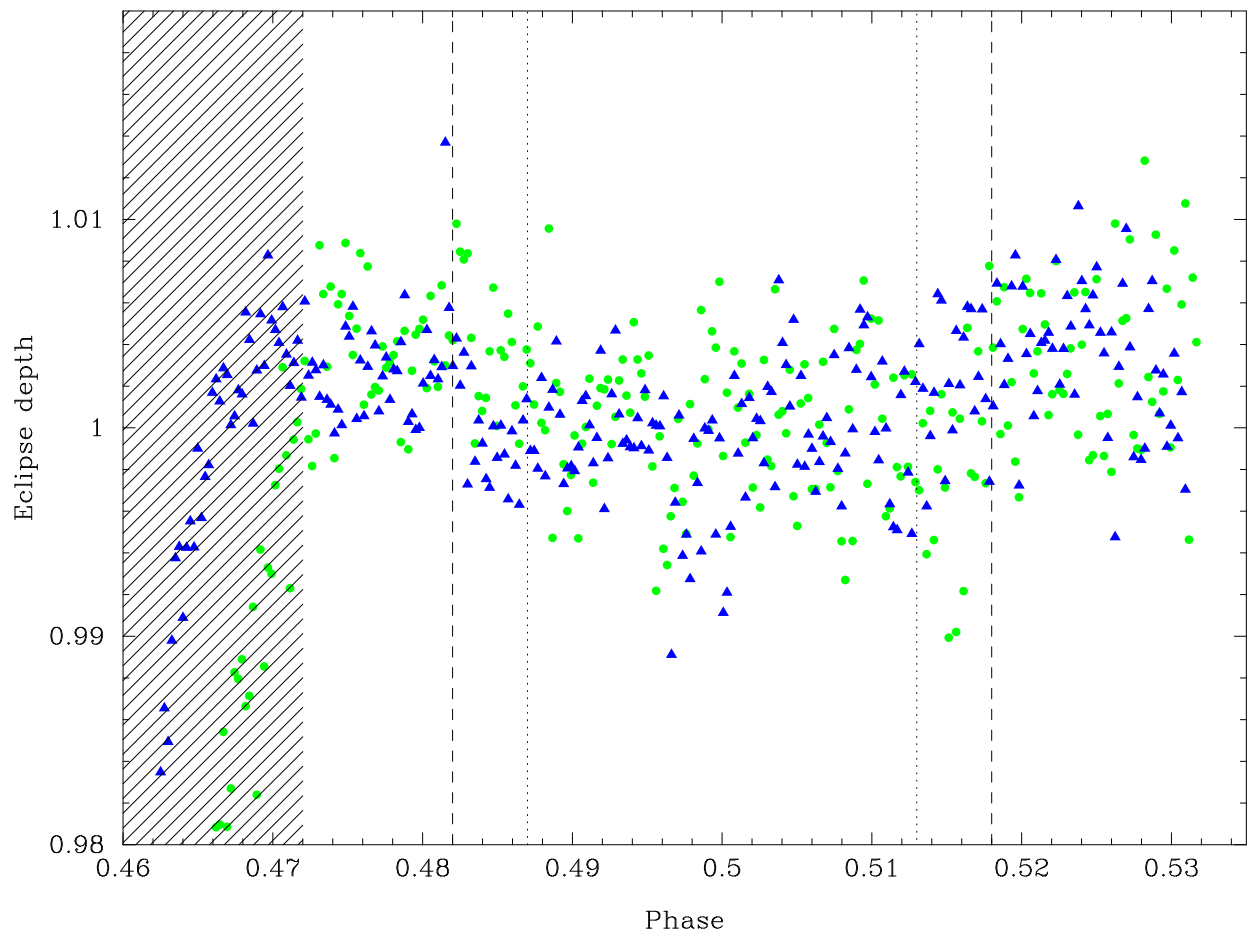}
\caption{The broad-band normalized light curves average between 7.6
and 14.2 $\mu$m.  The filled points are AOR 14817792, and the open
points are AOR 14818048.  The eclipse is clearly evident, as are the
ingress/egress transitions.  The rapid increase in the points at the
beginning of each AOR shows the accumulation (and subsequent
stabilization) of latent charge.  We have omitted the data affected by
latent charge (indicated by the shaded region) from our analysis, and
we find the average eclipse depth is 0.00315 $\pm$0.000315.
\label{fig:eclipseDepth}}
\end{center}
\end{figure}

\begin{figure}[h!]
\begin{center}
\epsscale{0.9} 
\includegraphics[angle=-90,scale=0.7]{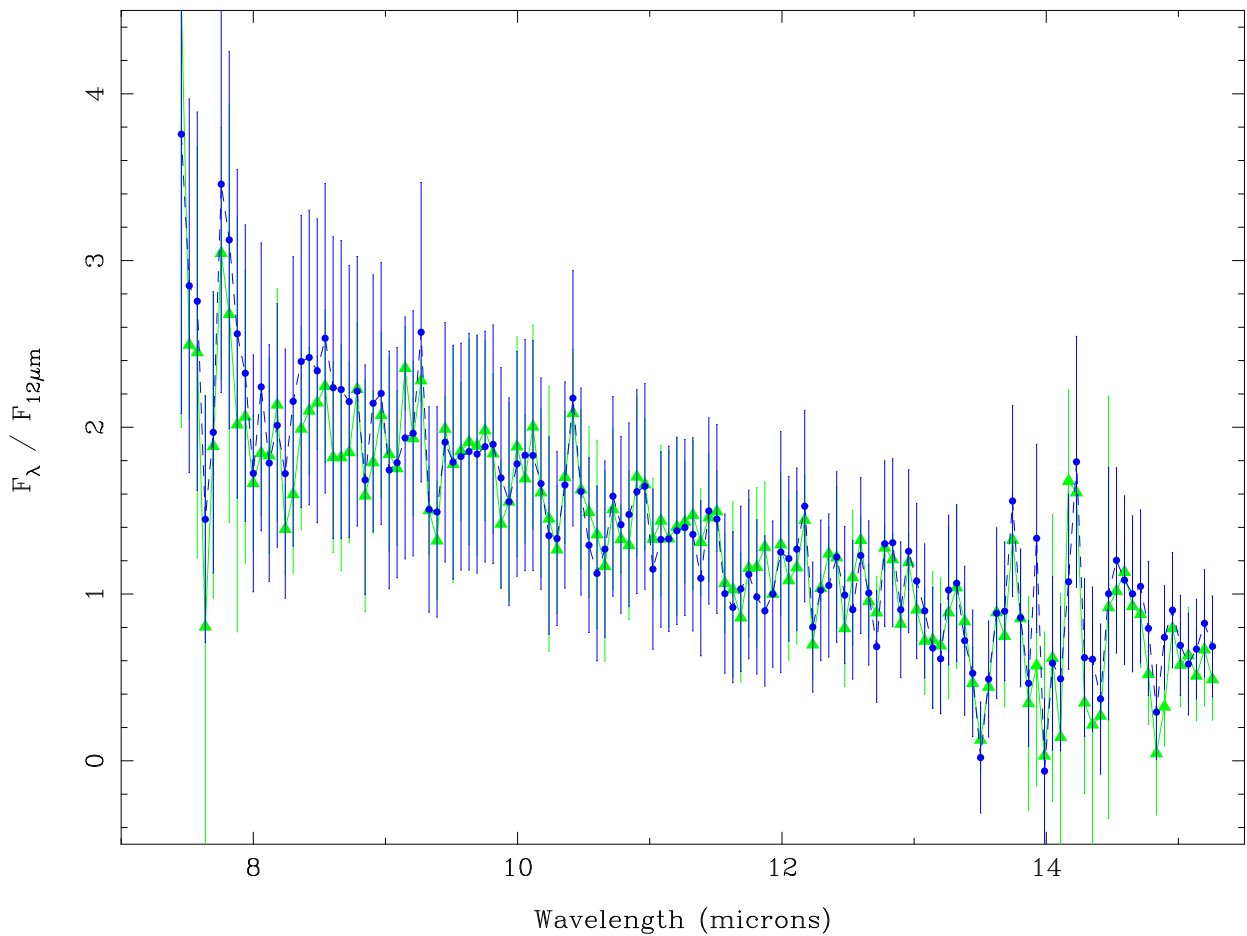}
\caption{The mid-infrared spectrum of HD 209458b derived using the
absolute method (circles) and the wavelength differential method
(triangles).  For both cases, the spectrum has been normalized to the
12 $\mu$m flux and scaled so that the average flux for both plots is
identical.  While the agreement is excellent over most of the
passband, the effect of small, uncorrected pointing errors causes the
differential spectrum to lie slightly below the absolute spectrum at
some wavelenghts between 7.5 and 9 $\mu$m.
\label{fig:specCompare}}
\end{center}
\end{figure}

\begin{figure}[h!]
\begin{center}
\epsscale{0.9} 
\includegraphics[angle=-90,scale=0.7]{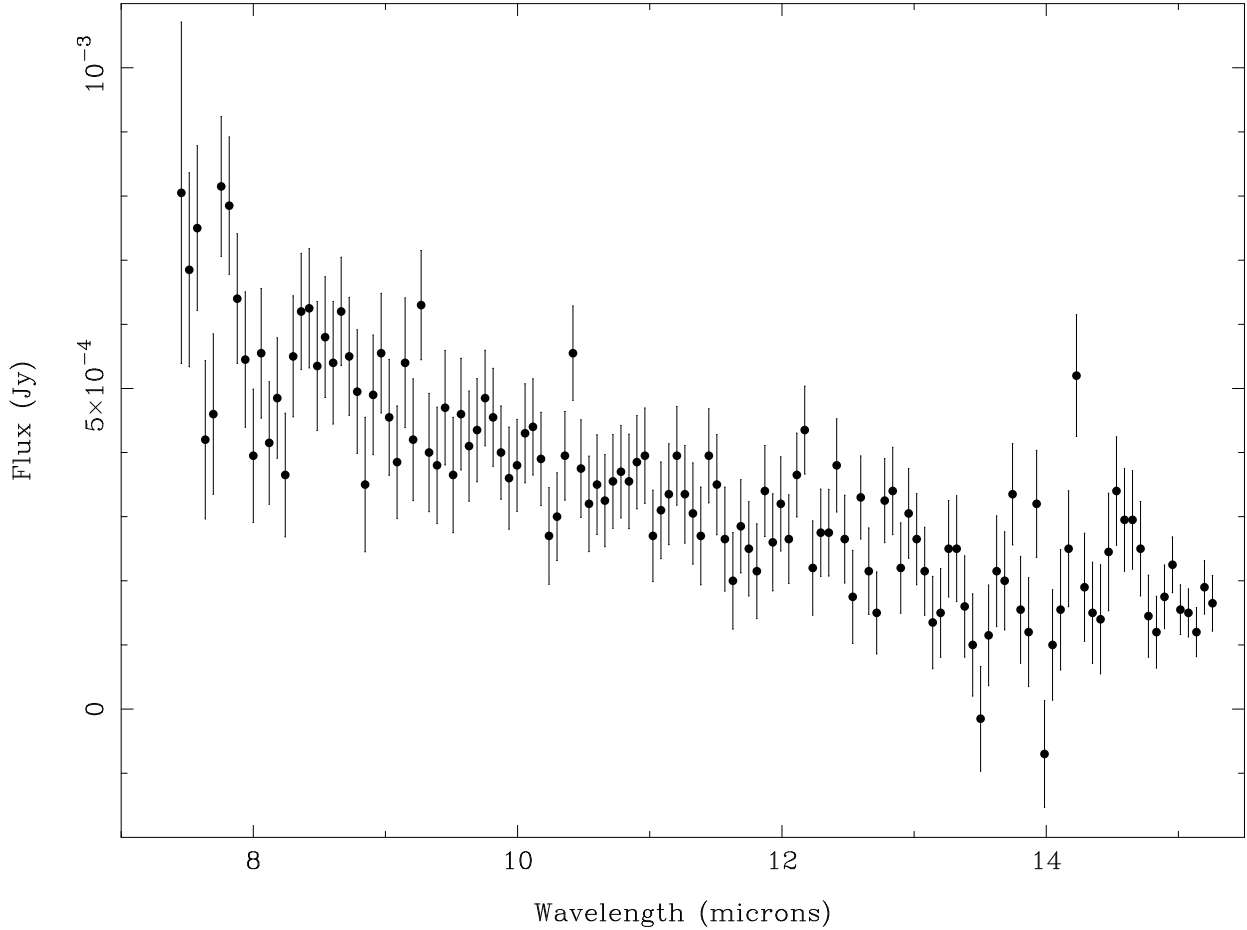}
\caption{The spectrum of HD 209458b is mostly smooth but contains some
modulation between 7.5 and 8.5 $\mu$m.  There is the suggestion of a
spectral feature at about 14.5 $\mu$m; however, the SNR in this region
of the spectrum is low, and the location of the spectral baseline is
uncertain.
\label{fig:absoluteSpec}}
\end{center}
\end{figure}

\begin{figure}[h!]
\begin{center}
\includegraphics[angle=-90,scale=0.7]{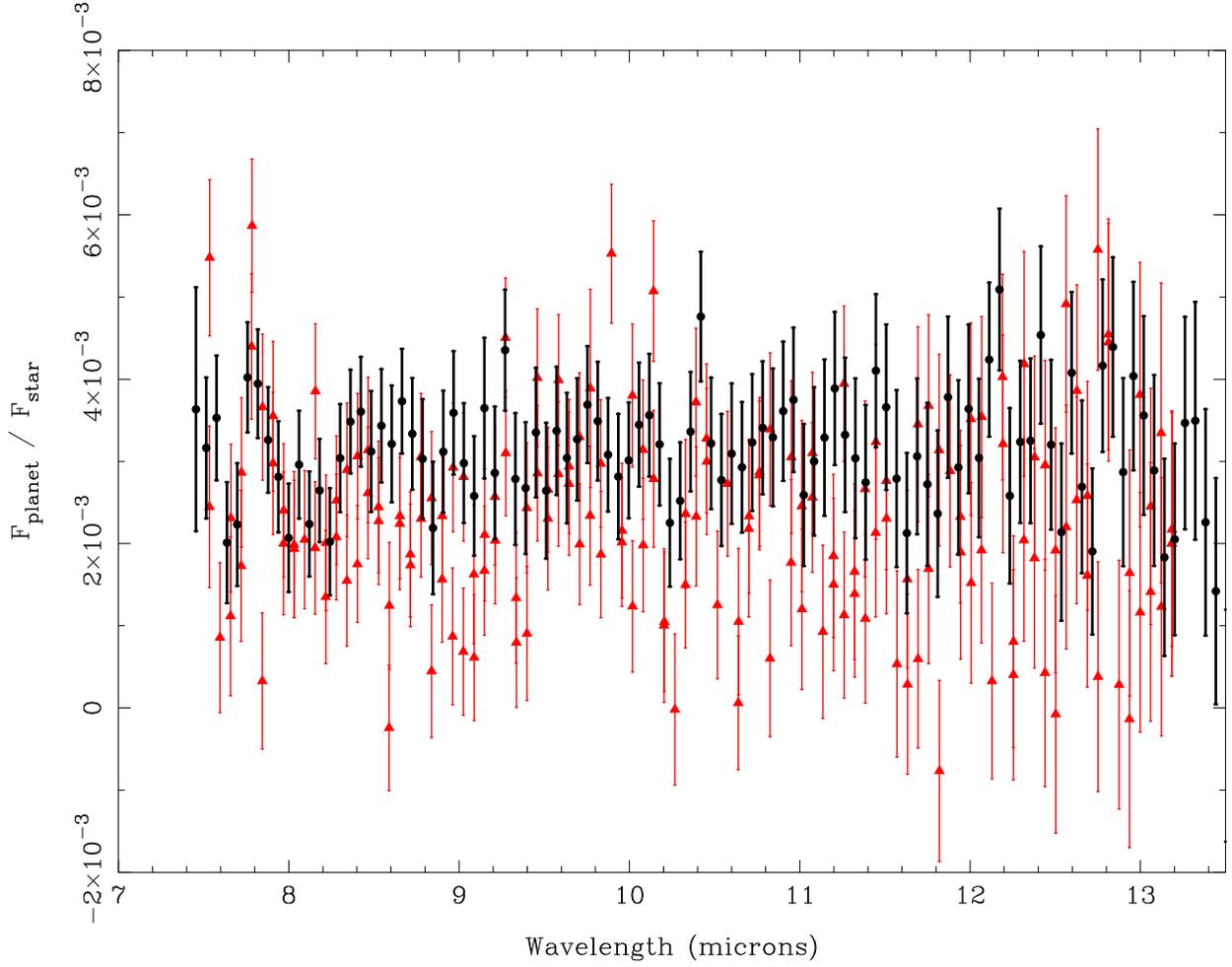}
\caption{The contrast spectrum for HD 209458b as determined by our
calibration method (black circles) and the method used by
\cite{richardson07} (red triangles).  Over the range of wavelengths where
the spectra can be directly compared, the internal scatter of our
spectrum is 2.3 times smaller than the previous result.  The methods
are in qualitative agreement, and both show evidence for modulation of
the spectrum between 7.5 and 8.5 $\mu$m.  This modulation includes a
broad feature seen as a local minima centered at approximately 8.1
$\mu$m, and a narrow feature that could be either a local minima at
approximately 7.57 $\mu$m or a local maxima centered at approximately
7.63 $\mu$m.
\label{fig:richardsonSpec}}
\end{center}
\end{figure}

\begin{figure}[h!]
\begin{center}
\includegraphics[angle=0,scale=0.7]{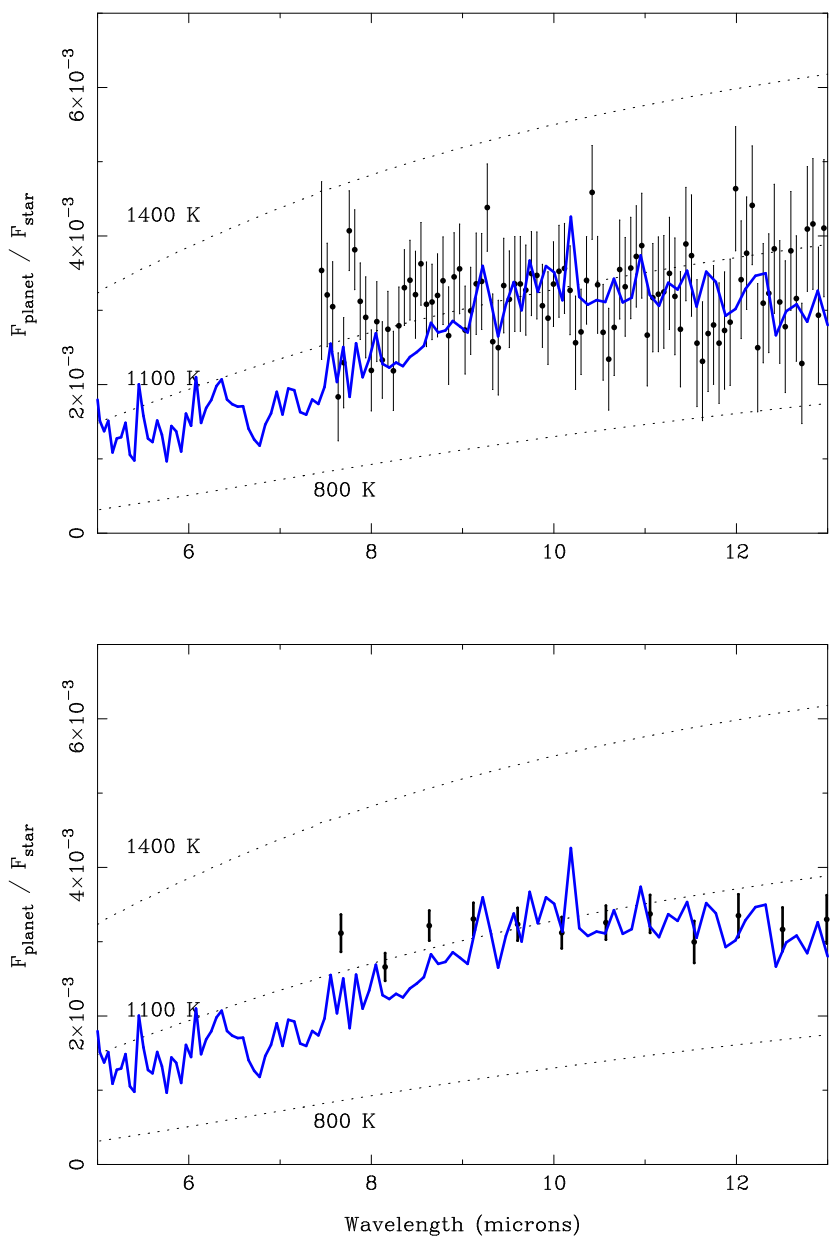}
\caption{The contrast spectrum for HD 209458b plotted together with a
model (Burrows et al. 2006) for the exoplanet emission.  The planet
emission is most consistent with models containing significant heat
redistribution. The departure of the measurement from the model in
the 8 $\mu$m region of the spectrum is significant and may place
constraints on models of the planet emission.  We have omitted
wavelengths beyond 13 $\mu$m from the contrast spectrum because these
data have lower SNR.
\label{fig:specFinal}}
\end{center}
\end{figure}

\begin{table}[h]
\begin{center}
Spatial Axis Fit Parameters \\
\begin{tabular}{|c|c|c|c|c|c|c|c|c|} \hline\hline
AOR/nod & $A_{\theta}$ & $\omega$ & $\phi_{\theta}$ & $x_{0}$ & $m_{x}$ & $A_{x}$ & $\phi_{x}$ \\ \hline
        & [rad/time] & [cycles/nod] & [radians] & [pix] & [pix/nod] & [pix] & [radians]\\ \hline \hline
AOR 7792/nod1 & 0.2699 & 2.9784 & 0.665 & -5.2497 & 0.0042  & 0.0167 & 0.818 \\
AOR 7792/nod2 &  ''    & 2.9785 & 0.765 &  5.835 & -0.0050  &  ''    & 0.918 \\
AOR 8048/nod1 &  ''    & 2.9787 & 0.085 & -3.871 & -0.0295  &  ''    & 0.238 \\
AOR 8048/nod2 &  ''    & 2.9785 & 0.135 &  6.25  & -0.0104  &  ''    & 0.288 \\ \hline
\end{tabular}
\label{tab:XpositionValues}
\caption{The parameters determined by fitting an elliptical pointing
error to the source position along the spatial axis of the
spectrometer slit.  $A_{\theta}$ is the amplitude of the angular
acceleration, $\omega$ is the angular frequency, $\phi_{\theta}$ is
the angular phase, $x_{0}$ is the constant offset in the x position,
$m_{x}$ is the linear change of position with time in the spatial
axis, $A_{x}$ is the amplitude of the $x$ periodic function, and
$\phi_{x}$, is the phase in the spatial axis.  }
\end{center}
\end{table}

\begin{table}[h]
\begin{center}
Spectral Axis Fit Parameters \\
\begin{tabular}{|c|c|c|c|c|} \hline\hline
AOR/nod & $y_{0}$ & $m_{y}$ & $A_{y}$ & $\phi_{y}$ \\ \hline 
        & [pixels] & [pixels/nod] & [pixels] & [phase] \\ \hline \hline 
AOR 14817792/nod1 & 0.2149 & -0.0242 & 0.0145 & 0.8474 \\ 
AOR 14817792/nod2 & 0.3003 & -0.0276 & `` &  `` \\
AOR 14818048/nod1 & 0.1175 & -0.0202 & `` &  `` \\
AOR 14818048/nod2 & 0.2023 & -0.0403 & `` &  `` \\ \hline
\end{tabular}
\label{tab:YpositionValues}
\caption{The parameters for the telescope pointing model in the
cross-slit (dispersion) axis: initial offset, $y_{0}$; linear drift,
$m_{y}$; periodic amplitude, $A_{y}$; and phase, $\phi_{y}$.  These
values were determined by requiring self-similarity of the spectrum
during the ``star+planet'' and ``star only'' intervals of the spectral
time series.  }
\end{center}
\end{table}

\begin{table}[h]
\begin{center}
Planet/Star Contrast Spectrum \\
\begin{tabular}{|c|c|c|} \hline\hline
wavelength ($\mu$) & contrast & error \\ \hline \hline
7.67 & 0.0031 & 0.00025 \\
8.15 & 0.0027 & 0.00019 \\
8.63 & 0.0032 & 0.00020 \\
9.12 & 0.0033 & 0.00022 \\
9.60 & 0.0032 & 0.00022 \\
10.09 & 0.0031 & 0.00021 \\
10.57 & 0.0033 & 0.00023 \\
11.05 & 0.0034 & 0.00026 \\
11.54 & 0.0030 & 0.00028 \\
12.02 & 0.0033 & 0.00029 \\
12.51 & 0.0032 & 0.00029 \\
12.99 & 0.0033 & 0.00033 \\
13.47 & 0.0025 & 0.00040 \\
13.96 & 0.0029 & 0.00046 \\
14.44 & 0.0040 & 0.00049 \\ \hline
\end{tabular}
\label{tab:spectrum}
\caption{Data and 1-$\sigma$ error for the contrast spectrum of HD
209458b shown in the lower panel of Fig. 13.}
\end{center}
\end{table}

\end{document}